\documentclass[sigconf,nonacm,authorversion]{acmart}

\usepackage{xcolor}
\usepackage[utf8]{inputenc}
\usepackage{amsmath}
\usepackage{graphicx} 
\usepackage{siunitx}
\usepackage{array}
\usepackage[frozencache]{minted}
\usepackage{soul}
\usepackage{tikz}
\usetikzlibrary{decorations.pathreplacing}
\usepackage{float}
\usepackage{algorithm}
\usepackage{algpseudocode}
\usepackage{flushend}

\usepackage{enumitem}
\usepackage{balance}
\usepackage{hyperref}
\usepackage{url}
\usepackage{xspace}
\usepackage{tcolorbox}
\usepackage{multirow}
\usetikzlibrary{shapes}
\usetikzlibrary{patterns, fit}
\usepackage{adjustbox}
\usepackage{hyperref}
\usepackage{pifont}
\setminted[java]{
xleftmargin=15pt,
framesep=2mm,
linenos=false,
breaklines=true,
tabsize=2,
encoding=utf8,
frame=none,
fontsize=\scriptsize
}

\makeatletter
\newcommand{\ostar}{\mathbin{\mathpalette\make@circled\star}}
\newcommand{\make@circled}[2]{%
  \ooalign{$\m@th#1\smallbigcirc{#1}$\cr\hidewidth$\m@th#1#2$\hidewidth\cr}%
}
\newcommand{\smallbigcirc}[1]{%
  \vcenter{\hbox{\scalebox{0.97778}{$\m@th#1\bigcirc$}}}%
}

\newcolumntype{g}{>{\columncolor{gray!30}}l}
\newcolumntype{h}{>{\columncolor{gray!30}}c}

\newcommand{\dataset}[0]{\textsc{AndroLibZoo}\xspace}

\newcommand{\az}[0]{\textsc{AndroZoo}\xspace}

\newcommand{\libsFromMaven}[0]{\num{69316}\xspace}
\newcommand{\libsFromGoogle}[0]{\num{235}\xspace}
\newcommand{\sumWithGoogle}[0]{\num{69535}\xspace}
\newcommand{\transitiveDependencies}[0]{\num{15089}\xspace}
\newcommand{\sumWithTransitiveDependencies}[0]{\num{69877}\xspace}
\newcommand{\libsOpenSourceProjects}[0]{\num{463}\xspace}
\newcommand{\sumWithLibsOpenSourceProjects}[0]{\num{69980}\xspace}
\newcommand{\libsOpenSourceProjectsRepositories}[0]{\num{6515}\xspace}
\newcommand{\sumWithLibsOpenSourceProjectsRepositories}[0]{\num{76007}\xspace}
\newcommand{\libsAfterRefinement}[0]{\num{34813}\xspace}

\newcommand{\dcircle}[1]{\ding{\numexpr171 + #1}}

\begin{document}

\title{\dataset: A Reliable Dataset of Libraries Based on Software Dependency Analysis}

\author{Jordan Samhi$^{\alpha}$, Tegawendé F. Bissyandé$^{\beta}$, Jacques Klein$^{\beta}$}
\affiliation{%
  \institution{$^{\alpha}$CISPA – Helmholtz Center for Information Security, jordan.samhi@cispa.de \\ $^{\beta}$SnT, University of Luxembourg, Luxembourg, \{tegawende.bissyande, jacques.klein\}@uni.lu}
  \country{}
}

\begin{abstract}
\textcolor{red}{This is the author's version of the work. It is posted here for your personal use. Not for redistribution. The definitive version is: https://doi.org/10.1145/3643991.3644866.}

Android app developers extensively employ code reuse, integrating many third-party libraries into their apps. 
While such integration is practical for developers, it can be challenging for static analyzers to achieve scalability and precision when libraries account for a large part of the code.
As a direct consequence, it is common practice in the literature to consider developer code only during static analysis  --with the assumption that the sought issues are in developer code rather than the libraries.
However, analysts need to distinguish between library and developer code.
Currently, many static analyses rely on white lists of libraries. 
However, these white lists are unreliable, inaccurate, and largely non-comprehensive.

In this paper, we propose a new approach to address the lack of comprehensive and automated solutions for the production of accurate and ``always up to date" sets of libraries.
First, we demonstrate the continued need for a white list of libraries.
Second, we propose an automated approach to produce an accurate and up-to-date set of third-party libraries in the form of a dataset called \dataset.
Our dataset, which we make available to the community, contains to date \libsAfterRefinement libraries and is meant to evolve.
\end{abstract}

\maketitle

\section{Introduction}
\label{sec:introduction}

Static analysis is a popular technique used in the literature to analyze Android apps, it analyzes app code without executing it. 
This approach is widely used to uncover security issues~\cite{10.1016/j.infsof.2017.04.001}. 
For example, researchers apply static analysis techniques to detect privacy leaks~\cite{10.1145/2666356.2594299,10.5555/2818754.2818791,10.1109/ICSE43902.2021.00126,10.1145/3510003.3512766,10.1145/2660267.2660357} and detect malicious code~\cite{10.1109/SP.2016.30,10.1145/3510003.3510135,10.1007/978-0-387-68768-1_4,10.1145/3574158}. 
However, most of these approaches need to differentiate between developer and library code to focus on the app's functionality, which is the relevant code for finding security problems and avoiding scalability issues (due to the widespread use of polymorphism in libraries and the over-approximation of static analyzers).
This is why static analyzers do not dive into the Android framework code during analyses (e.g., FlowDroid discards classes that are within the Android framework, cf. lines 64--69 in FlowDroid's SystemClassHandler class~\cite{flowdroidSystemClassHandler}).
Differentiating between developer and library code is, therefore, a crucial step for static analysis to be effective and more scalable~\cite{9286020,9542854}, as it allows analyzers to focus on the parts of the app that are most likely to contain security issues.

Furthermore, libraries can introduce noise for malware detection.
For example, Mudflow~\cite{10.5555/2818754.2818808} and DroidAPIMiner~\cite{10.1007/978-3-319-04283-1_6} show that discarding libraries in their analyses improves their malware detection performance. 
A reliable list of libraries is thus an important artifact for the research and analyst community. 

Previous studies have employed white lists 
to identify libraries in Android apps.
Chen et al.~\cite{10.1145/2568225.2568286} manually compiled a list of 73 package names from common libraries. 
Similarly, Grace et al.~\cite{10.1145/2185448.2185464} randomly selected apps from a dataset of \num{100000} apps that were manually screened to identify libraries.
With this approach, they created a list of 100 libraries. 
These lists are ad-hoc and incomplete, 
as they only contain 73 and 100 libraries, respectively.
Another method to build a white list of libraries has been proposed by Li et al.~\cite{10.1109/SANER.2016.52}.
Their approach involves using a large dataset of apps to identify candidate libraries. 
The process includes ranking all package names by frequency of appearance in apps and using heuristics to retain libraries.
However, though the approach is considered the \emph{state-of-the-art} white list of libraries in the literature~\cite{9713838},
the list provided is outdated, 
the method used to create the list relies on arbitrary heuristics, 
and the hypothesis to consider a package name as a library according to its occurrence leads to a high rate of false positives (if numerous versions of the same app are present, for instance: \num{20715} different versions of app ``com.slideme.sam.manager" are present in \az).
Hence, creating a comprehensive white list of libraries to discriminate the developer code from libraries accurately remains an open challenge.

In this work, we propose a novel approach to build the first extensive and precise, by construction, white list of libraries by mining software dependencies.
Contrary to the research literature, which often relies upon complex approaches~\cite{10.1145/3324884.3416582},
our method involves mining information from developer habits.
We propose to the research and analyst communities a dataset called \dataset, containing \libsAfterRefinement libraries.
\dataset is accurate by construction, i.e., it only contains libraries.
This dataset is meant to evolve and regularly incorporate new libraries added by third-party vendors.
\dataset aims to facilitate the work of static analysis analysts in terms of scalability and robustness.
We believe this dataset will be a valuable resource for static analysis, we encourage its use and expansion as new libraries become available.

Overall, this paper makes the following contributions:
\begin{itemize}[noitemsep,topsep=0pt]
    \item We show that white list of libraries are still needed.
    \item We propose an approach to build an accurate set of libraries.
    \item We build \dataset, the first version of the library dataset produced by our approach.
\end{itemize}

Our artifacts are available:
\begin{center}
\url{https://github.com/JordanSamhi/AndroLibZoo}
\url{https://zenodo.org/records/10072709}
\end{center}

\section{Motivation}
\label{sec:motivation}

In this section, we motivate our work with
a study to demonstrate the need for a white list of libraries.

Despite existing approaches' limitations, it is unclear whether white lists of libraries are still necessary in practice, mostly due to the usage of obfuscation. 
Hence, a legitimate question is:
\textbf{To what extent does obfuscation jeopardize the use of a white list in Android apps?}
To address this question, we conducted a motivating study examining the prevalence of package name obfuscation in Android apps.

\begin{figure}
    \centering
    \begin{adjustbox}{width=\columnwidth,center}
        \includegraphics[scale=1]{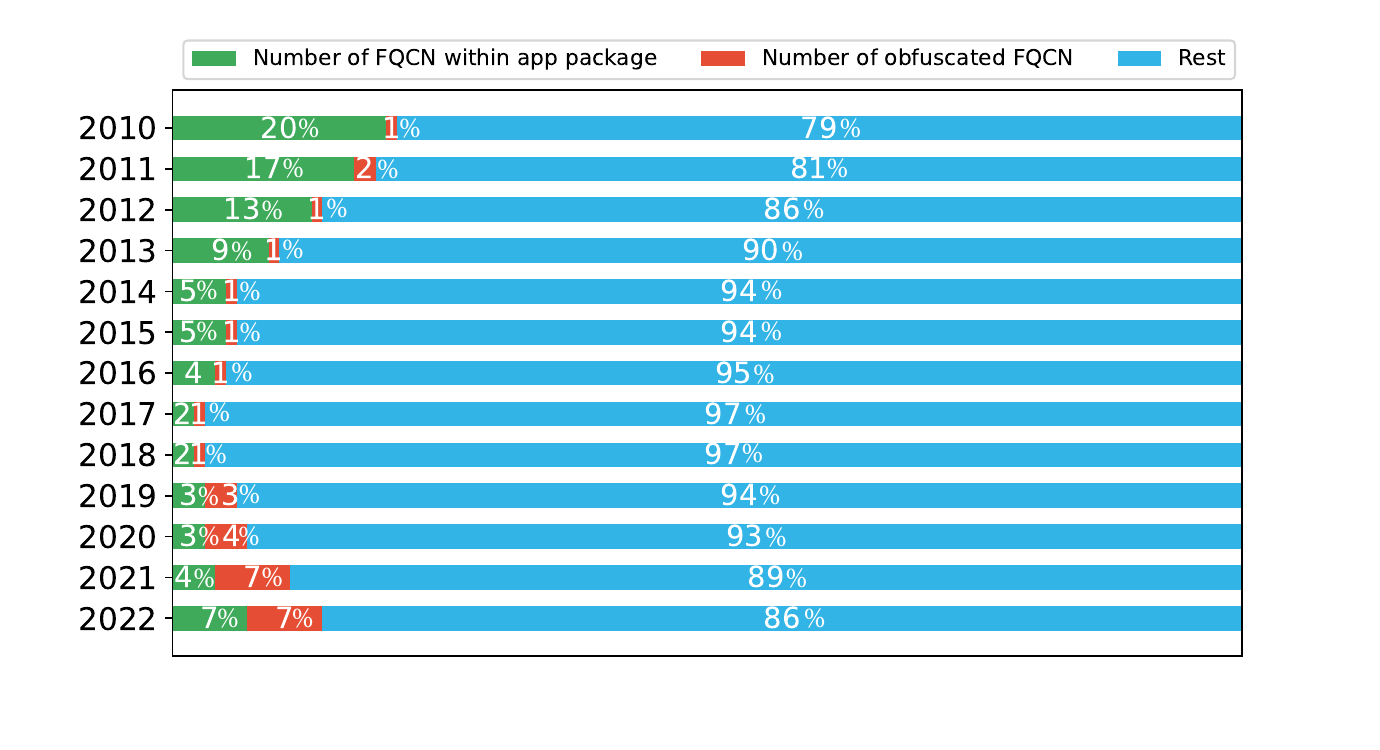}
    \end{adjustbox}
    \caption{Proportion of FQCNs within the apps' package and obfuscated FQCNs}
    \label{fig:fqcn_motivation}
\end{figure}

Let us first introduce name obfuscation, a technique used to remove package names and/or class names' meaning  to prevent or hinder reverse engineering and make malicious code detection more difficult. 
Typically, package and class names have meaningful names to make it easier for developers to understand the app structure and the code's purpose.
Obfuscation is both used to protect apps' intellectual property and to make malicious code harder to detect.
Most code obfuscators replace package or class names with simple letters from the alphabet~\cite{LI2019157,10.1109/SANER.2016.52}. 
For instance, the package name ``com.example.myapp.MyClass" could be obfuscated such as:
\dcircle{1} only the class name could be ofuscated: \texttt{com.example.myapp.a};
\dcircle{2} only the package name could be obfuscated: \texttt{a.b.c.MyClass};
\dcircle{3} both the package name and the class name can be obfuscated: \texttt{a.b.c.d.a}; or
\dcircle{4} the package name can be removed: \texttt{a}.

To conduct this study, we randomly selected \num{10000} apps per year from the Androzoo~\cite{10.1145/2901739.2903508} dataset for each year from 2010 to 2022 and checked whether any of the Fully Qualified Class Names (FQCNs) in these apps either:
\dcircle{1} started with a letter of the alphabet (e.g., \texttt{a.b.*}, or \texttt{g.u.*}); or
\dcircle{2} its class name is a single letter of the alphabet (e.g., \texttt{com.example.a} or \texttt{f.w.i}).
This allows us to cover all the cases cited above.

The results of this study are shown in Figure~\ref{fig:fqcn_motivation}.
This figure presents three categories of FQCNs:
\dcircle{1} the number of fully qualified class names that are within the package name of the app, i.e., the one declared in the manifest (e.g., if the package name of the app is \texttt{org.example} and FQCN is \texttt{org.example.MyClass});
\dcircle{2} the number of FQCNs that have been obfuscated, as defined previously;
\dcircle{3} the number of all remaining FQCNs, i.e., FQCNs that are not within the package name of the apps nor in an obfuscated package name.

This study shows that a low percentage of classes are in the app package (i.e., package with the app package name).
But more importantly, the percentage of obfuscated FQCNs in the apps is even lower, even if this percentage appears to be growing over time.
The largest percentage is made up of FQCNs that are neither within the package name of the app nor obfuscated. 
These results suggest that the use of name obfuscation techniques is not yet widespread enough to render white lists unnecessary for static analysis of Android apps.
Therefore, it appears that white lists 
are still useful for differentiating these FQCNs from developer and library code.

\section{Dataset Construction Methodology}
\label{sec:methodology}

This section presents our methodology to build \dataset.

\begin{figure}
    \centering
    \begin{adjustbox}{width=\columnwidth,center}
        \input{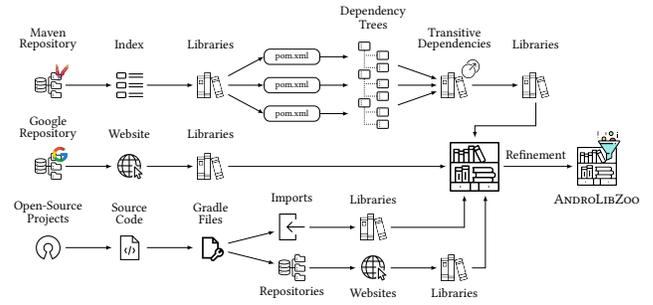}
    \end{adjustbox}
    \caption{Overview of our methodology to construct \dataset, a collection of third-party libraries.}
    \label{fig:overview}
\end{figure}

\noindent
\textbf{Hypotheses:}
Our hypothesis is that:
\dcircle{1} the majority of Android app developers rely on Android Studio and Gradle to build their apps;
\dcircle{2} that Gradle in turn relies on Maven and Google to retrieve libraries~\cite{gradle_maven, gradle_google}; and
\dcircle{3} open-source Android apps can bring valuable information about the library used in apps, thanks to the availability of configuration files.

Based on these hypotheses, we set up a straightforward approach that can be seen in Figure~\ref{fig:overview}.
Our approach involves two main steps:
\dcircle{1} extracting libraries from the Maven and Google's Android library repositories; and 
\dcircle{2} extracting libraries used in open-source projects.
Maven is a widely used tool for managing dependencies and building Java-based projects, and the Maven repository is a central location where developers can find and share libraries. 
Google's Android library repository is a collection of Android libraries available through the Google Maven repository.
In the following sections, we explain how we proceed to build \dataset.

\subsection{Mining repositories}

This section details how we obtained the first portion of our dataset by leveraging public library repositories.
Our process is divided into two steps:
\dcircle{1} obtaining package names from the repositories.
\dcircle{2} gathering transitive dependencies from the  package names.

\subsubsection{Package Name Collection}

This section explains how we collect package names from Maven and Google repositories.

\noindent
\textbf{Maven.} 
First, we download the Maven index file (updated weekly) and the Maven Indexer Command Line Interface (CLI) tool which are available publicly.
The Maven index is a file created using the Lucene library~\cite{bialecki2012apache} and contains metadata about the artifacts available in the Maven repository. 
The Maven Indexer CLI is a tool used to extract this file. 
Once the Maven index file is extracted, we use a Java program of our own, called SearchLuceneIndex, to extract all the \emph{groupIds} from the index.
At the time of writing, our list from the Maven repository contains a total of \libsFromMaven entries.

\noindent
\textbf{Google.}
We could extract a total of \libsFromGoogle libraries by accessing the Maven Google repository.
Once we have obtained the package names from both the Maven repository and the Maven Google repository, we merge the two lists into a single list, eliminating any duplicate entries. This results in a list of \sumWithGoogle package names.

\subsubsection{Transitive Dependency Extraction}

Libraries often rely on other libraries known as \emph{transitive dependencies}. 
For example, the \texttt{com.android.tools.sdk-common} library in version 22.9.0 has 12 transitive dependencies. 
Figure~\ref{code:tree} shows the dependency tree of this library.
To create a comprehensive and up-to-date set of libraries, it is important to consider these transitive dependencies. 

\begin{figure}
\begin{minted}{java}
com.android.tools:sdk-common:22.9.0
\- com.android.tools:sdklib:22.9.0
   +- com.android.tools:dvlib:22.9.0
   |  \- com.android.tools:common:22.9.0
   |     \- com.google.guava:guava:15.0
   +- org.apache.httpcomponents:httpclient:4.1.1
   |  +- org.apache.httpcomponents:httpcore:4.1
   |  +- commons-logging:commons-logging:1.1.1
   |  \- commons-codec:commons-codec:1.4
   +- com.android.tools.layoutlib:layoutlib-api:22.9.0
   |  \- net.sf.kxml:kxml2:2.3.0
   +- org.apache.httpcomponents:httpmime:4.1
   \- org.apache.commons:commons-compress:1.8.1
\end{minted}
      \caption{Dependency tree of the com.android.tools.sdk-common library version 22.9.0.}
    \label{code:tree}
\end{figure}

We used the \emph{mvn} utility (from Maven) to build a given project's dependency tree.
However, since our dataset is made of libraries (i.e., package names) and not actual development projects, getting the dependency trees from the libraries is not trivial. 
We implemented the following strategy.
First, from the maven index, we obtained all artifacts in the form of the artifactIds (e.g., com.example) concatenated with groupIds (e.g., MyLibrary) and versions (e.g., 1.0) to get the real library names, resulting in a list of \num{10069375} unique libraries.
We considered all available versions for each library because transitive dependencies can differ from one version of a library to another. 
Then, for each library, we generated a \texttt{pom.xml} file with a dummy project and the library as a dependency of the project.
We then launched the \texttt{mvn dependency:tree} command to get the list of all transitive dependencies of the library.
Obtaining transitive dependencies for all libraries in our dataset is computationally intensive. 
With a total of \num{10069375} libraries, extracting transitive dependencies for each library would have taken an unreasonable amount of time. 
Therefore, we set a timeout of 30 seconds for each call to the \texttt{mvn dependency:tree} command.
Even with this timeout, \num{9312664} (i.e., 92.5\%) dependency trees were successfully built, and the process still required 21 days of computation using 70 instances in parallel on a Debian server with an AMD EPYC 7552 48-Core Processor CPU with 96 cores and 630GB of RAM.
Then, we parsed the dependency trees generated and built a list of \transitiveDependencies additional libraries with any duplicate removed.
It should be noted that this number (i.e., \transitiveDependencies) was reached long before (after only 10 days) the \num{9312664} dependency trees were built.
Finally, we merged this list with the one generated previously, and we removed any duplicates, resulting in a final list of \sumWithTransitiveDependencies package names.

\subsection{Mining open-source Android projects}

Open-source projects provide valuable information about the libraries, relying on various third-party libraries to provide specific features and functionality. 
In the following, we describe the process for extracting lists of libraries from open-source projects.

We first obtained a list of open-source Android projects by downloading all apps in \az that were collected from the F-Droid repository.
At the time of writing, this represented \num{4464} apps. 
We then used the F-Droid website to crawl the source code links for these apps and attempted to clone the repositories.
Out of the 4464 apps, we could successfully clone 3425 projects (some reasons why some repositories cannot be cloned involve repositories that do not exist anymore, or the "git clone" command cannot work for, e.g., svn repositories, our prototype currently ignores non-git repositories).
Next, we searched the \texttt{build.gradle} files available in these projects and extracted information about the libraries used.
Specifically, we searched for the Gradle commands \texttt{implementation}, \texttt{classpath}, and \texttt{compile}, commonly used to declare dependencies. 
After parsing the \texttt{build.gradle} files and extracting the relevant information, we obtained a list of \libsOpenSourceProjects unique libraries, which brings our total count of libraries to \sumWithLibsOpenSourceProjects after deduplicating.

To obtain a list of libraries as comprehensive as possible, it is not enough to consider libraries explicitly imported in the \texttt{build.gradle} files of the app projects.
It is also necessary to consider library repositories (other than the default Maven and Google) declared in the \texttt{build.gradle} files and use this information to search for additional libraries.
This process led to two additional repositories: jcenter and gradlePluginPortal. 
However, we found jcenter is deprecated and all its libraries are now in the Maven repository~\cite{jcenter},  
which we have already mined. 
The gradlePluginPortal website proved to be a valuable resource. 
We crawled the gradlePluginPortal website, comprising 656 pages at the time of writing, and obtained an additional list of all package names available.
This led to the discovery of \libsOpenSourceProjectsRepositories libraries, which we added to our list after removing potential duplicates and led to a total of \sumWithLibsOpenSourceProjectsRepositories libraries.

\subsection{Refinement}
We have designed a refinement process to remove unnecessary package names from the list of third-party libraries. This process helps ensure that the list is as concise as possible.
For example, if the list contains the package names ``com.example.subpackage" and ``com.example", we would only keep ``com.example" because this would be sufficient for identifying ``com.example.subpackage.My\-Class" as a library in an app.
This is done by iterating over the list of package names and checking whether any element $e_1$ in the list starts with another element $e_2$ in the same list. 
If this is the case, the element $e_1$ is removed from the list. This process is repeated until no more changes are made to the list.
Overall, our refinement process yields a list of \libsAfterRefinement package names by discarding \num{41194} of them (i.e., \sumWithLibsOpenSourceProjectsRepositories $-$ \libsAfterRefinement).
Table~\ref{table:steps} shows the details of all the steps of our methodology with the number of libraries collected.

\begin{table}
    \setlength{\tabcolsep}{1pt}
    \centering
    \caption{\# of libraries after each step (OS = open-source)}
    \begin{adjustbox}{width=.8\columnwidth,center}
        \begin{tabular}{lccccccccc}
            \hline
            Source &  &  \\ \hline
            Maven & \libsFromMaven &  \\ 
            Google & +\libsFromGoogle &$\rightarrow$& \sumWithGoogle \\ 
            Transitive Dep. && &+\transitiveDependencies & $\rightarrow$ & \sumWithTransitiveDependencies && \\ 
            OS imports & &&&&+\libsOpenSourceProjects & $\rightarrow$ & \sumWithLibsOpenSourceProjects &\\ 
            gradlePluginPortal &&& &&&&+\libsOpenSourceProjectsRepositories & $\rightarrow$ &\sumWithLibsOpenSourceProjectsRepositories \\ \hline
            After Refinement &&&&&&&& & \libsAfterRefinement \\ \hline
        \end{tabular}
    \end{adjustbox} 
    \medskip
    \textit{each resulting number is given after removing duplicates}
    \label{table:steps}
\end{table}

\section{Dataset Description}
\label{sec:dataset}

In this section, we describe our dataset. 
\dataset contains \num{1210} unique roots.
A root refers to the first element in a package name. 
For example, in "com.example.mypackage", "com" is the root.

We also collected data on the number of fields in the apps' package names in our dataset. 
A field refers to a dot-separated element in a package name.
For example, in "com.example.mypackage", there are three fields: "com", "example", and "mypackage".
Results are visible in Table~\ref{table:num_of_fields}.
There are 732 package names with only one field, \num{17172} package names with two fields, etc.

\begin{table}
    \centering
    \caption{Number of package names per field in \dataset}
    \begin{adjustbox}{width=.8\columnwidth,center}
        \begin{tabular}{lr|lr}
            \hline
            Fields & Count & Fields & Count \\ \hline
            with 1 field & \num{732} & with 6 fields & \num{134}\\ 
            with 2 fields & \num{17172} & with 7 fields & \num{40}\\ 
            with 3 fields & \num{13086} & with 8 fields & \num{26}\\ 
            with 4 fields & \num{2979} & with 9 fields & \num{8}\\
            with 5 fields & \num{634} & with 10 fields & \num{2}\\
            \hline \hline
            \multicolumn{2}{l|}{Total} & \multicolumn{2}{r}{\libsAfterRefinement}
        \end{tabular}
    \end{adjustbox}
    \label{table:num_of_fields}
\end{table}

There are significantly fewer package names with four or more fields. 
The number of package names with one field is relatively low compared to the others.
This suggests that many package names in the dataset follow a standard naming convention with a domain name followed by one or more subpackages.
The presence of package names with four or more fields may indicate the use of more complex or specialized naming conventions.

Table~\ref{table:top_ten} presents the top 10 most frequent roots and the top 10 most frequent fields. 
In the first two columns, we see that ``com" is by far the most frequent root, with more than \num{13000} occurrences. 
The second most frequent root is ``org", with \num{5450} occurrences. 
In the second two columns, representing the most frequent fields, including the roots, we see that ``com" field is still the most frequent. 
The second two columns do not differ much from the first two columns, except for the field ``gradle" that now appears.
This could indicate that Android libraries are prevalent (often built with gradle).
It is confirmed in the last two columns, representing the most frequent fields without the roots, in which we see that ``gradle``, and ``android`` fields are the most frequent, with 680 and 443 occurrences respectively. 
After ``gradle" and ``android", the third most frequent field is ``maven", with 352 occurrences.
We see a shift in the most prevalent fields. 
Instead of roots, we now see fields such as ``sdk", ``maven", ``plugin(s)", ``api", and ``tools".
This may be indicative of the types of libraries.
Overall, the results suggest that most libraries are from the ``com" domain and Android libraries are well represented.

\begin{table}
    \centering
    \caption{Top 10 roots and fields present in \dataset}
    \begin{adjustbox}{width=\columnwidth,center}
        \begin{tabular}{c|c|c|c|c|c}
            \hline
            \multicolumn{2}{c|}{\textbf{Top 10 roots}} & \multicolumn{2}{c|}{\textbf{Top 10 most used fields}} & \multicolumn{2}{c}{\textbf{Top 10 most used fields w/o roots}}\\ \hline
            \textbf{Root} & \textbf{Count} & \textbf{Field} & \textbf{Count} & \textbf{Field} & \textbf{Count} \\ \hline \hline
            com & \num{13514} & com & \num{13844} & gradle & \num{680} \\ \hline
            org & \num{5450} & org & \num{5519} & android & \num{443} \\ \hline
            io & \num{2630} & io & \num{2646} & maven & \num{352} \\ \hline
            net & \num{1651} & net & \num{1704} & com & \num{330} \\ \hline
            de & \num{1287} & de & \num{1288} & plugins & \num{269} \\ \hline
            cn & \num{784} & cn & \num{784} & sdk & \num{267} \\ \hline
            dev & \num{562} & gradle & \num{697} & plugin & \num{258} \\ \hline
            me & \num{548} & dev & \num{575} & co & \num{244} \\ \hline
            eu & \num{329} & me & \num{557} & tools & \num{164} \\ \hline
            se & \num{304} & co & \num{460} & api & \num{153} \\ \hline
        \end{tabular}
    \end{adjustbox}
    \label{table:top_ten}
\end{table}

\section{Future Research Questions}
\label{sec:future_research_question}

This dataset opens avenues for several analyses.
This section draws three research questions that this dataset could be used to address.

\noindent
\textbf{Research Question 1:}
\textit{How does \dataset compare with state-of-the-
art approaches?} With this RQ, \dataset can be compared to existing techniques such as Li et al.~\cite{10.1109/SANER.2016.52} or Ma et al.~\cite{10.1145/2889160.2889178}. Researchers can assess their comprehensiveness and precision.

\noindent
\textbf{Research Question 2:}
\textit{To what extent can \dataset be useful and
effective for static analysis?}
This RQ would evaluate the importance of \dataset. 
Researchers can extract all packages from a dataset of apps and check the number of packages filtered using \dataset.
This quantity can then be compared with the same approach using existing lists or more straightforward approaches, such as only considering the apps' package names to filter libraries.

\noindent
\textbf{Research Question 3:}
\textit{Can \dataset improve the analysis performance
of existing static analysis tools?}
With this RQ, \dataset can be used within existing static analyzers to check whether their performances, in terms of scalability and precision, is improved or not.

\section{Limitations}
\label{sec:limitations}

One limitation of our work is that we did not consider all potential sources of third-party libraries publicly available. 
While we extracted libraries from Maven and Google's repositories, as well as open-source Android projects, there may be other sources of libraries.
This limitation is mitigated by the fact that given our hypothesis, we believe that the sources of libraries we relied on are representative of how developers build apps in general.

Another limitation is that we only considered a subset of open-source Android projects when extracting libraries. 
While we used the \az and the F-Droid datasets as sources of open-source projects, there are likely additional open-source projects available.

Another limitation of our work is that our list of libraries is only designed to match non-obfuscated libraries.
Our study has shown that this limitation is also mitigated (cf. Figure~\ref{sec:motivation}) since the majority of package names in Android apps are not obfuscated.
Besides, the detection of obfuscated libraries is another research direction that is actively being explored by the literature~\cite{10.1145/2976749.2978333,8543426,8330204}.
\section{Conclusion}
\label{sec:conclusion}

 In this paper, we presented an approach for automatically generating an accurate and up-to-date white list of third-party libraries 
that can serve the research and practitioner communities. 
Our dataset, \dataset, contains \libsAfterRefinement package names which, by construction, only represent libraries, and is meant to evolve.

\newpage

\bibliographystyle{acm}
\bibliography{bib}

\end{document}